# FemtoSats for Exploring Permanently Shadowed Regions on the Moon

Álvaro Díaz-Flores
SpaceTREx Laboratory
University of Arizona
1130 N Mountain Ave
Tucson, AZ 85721
adiazflores@email.arizona.edu

José Fernández
SpaceTREx Laboratory
University of Arizona
1130 N Mountain Ave
Tucson, AZ 85721
josef@email.arizona.edu

Leonard Vance
SpaceTREx Laboratory
University of Arizona
1130 N Mountain Ave
Tucson, AZ 85721
ldvance@email.arizona.edu

Himangshu Kalita
SpaceTREx Laboratory
University of Arizona
1130 N Mountain Ave
Tucson, AZ 85721
hkalita@email.arizona.edu

Jekan Thangavelautham
SpaceTREx Laboratory
University of Arizona
1130 N Mountain Ave
Tucson, AZ 85721
jekan@arizona.edu

**The recent, rapid advancement in space exploration is thanks to the accelerated miniaturization of electronics components on a spacecraft that is reducing the mass, volume and cost of satellites. Yet, access to space remains a distant dream as there is growing complexity in what is required of satellites and increasing space traffic. Interplanetary exploration is even harder and has limited possibilities for low cost mission. All of these factors make even CubeSats, the entry-level standard too expensive for most and therefore a better way needs to be found.**

**The proposed solution in this report is a low-mass, low-cost, disposable solution that exploits the latest advances in electronics and is relatively easy to integrate: FemtoSats. FemtoSats are sub-100-gram spacecraft. The FemtoSat concept is based on launching a swarm where the main tasks are divided between the members of the swarm. This means that if one fails the swarm can take its place and therefore substitute it without risking the whole mission.**

**In this paper we explore the utility of FemtoSats to perform first exploration and mapping of a Lunar PSR. This concept was recognized as finalist for the NASA BIG Competition in 2020. This is an example of a high-risk, high-reward mission where losing one FemtoSat does not mean the mission is in danger as it happens with regular satellite missions. The work developed here consists of making a conceptual generic design for FemtoSats and apply it to the exploration of the PSR's. The work results in the development of laboratory prototypes and simulations. A key approach we explore using FemtoSats is the use and implementation of LASER power beaming, to power and keep-alive the FemtoSats in a cold-dark region such as the PSRs. Within the PSR, we want to prove that batteries can be charged through a laser beam. Currently, the FemtoSat design for the PSR is being tested and first results look promising. Advances made on this project will take FemtoSat technology one major step closer towards flight readiness and operations and eventually towards applications on ambitious next-generation science exploration missions.**

## Introduction

At current technological trends, we are seeing electronics, sensors and actuators getting more compact, reliable and have increased energy conversion efficiency. This has been driven by rapid advancement and commercial appetite for smartphones, personal computing devices and for smart devices. Advances in these terrestrial areas have now crossed over to space applications. Thanks to the high-reliability of Commercial Off The Shelf (COTS), it is possible to use these components in space onboard entry-level, low-cost or disposable spacecraft such as CubeSats [1]. These COTS components, with sufficient redundancy could survive long-enough on short high-risk, high-reward mission in extreme environments.

The ever increasing popularity of CubeSats and small satellites have been studied from an economic point of view and they were seen as a disruptive innovation. According to [2], for small satellites to overtake the satellites market several conditions need to be met:
- New entrants must be identified in space industry
- Satellite miniaturization must meet the conditions of a disruptive innovation
- A complementary innovation must appear to boost satellite miniaturization as a disruptive innovation
- A path dependency must be observed in the existing firms
- Existing firms must implement open innovation
- New entrants must in turn practice innovation

At the time of this study most of the answers to this question supported the idea of a disruptive innovation but lacked a complementary innovation: launch prices. However, SpaceX advances in reusable rockets promises to greatly reduce launching prices which could be considered a complementary innovation. Therefore, small satellites are steadily acquiring the characteristics of disruptive innovations.

The current trend towards small satellites are CubeSats, thanks to miniaturization and COTS (Components Of The Shelf). These allowed engineers to introduce several devices in a single satellite or put a few small satellites in communication to perform a greater mission. However, this concept is still expensive since a typical 3U CubeSats mass is 4 kg. Current launching costs are $60,000 per kilogram according to SpaceX and typical CubeSats parts costs have ranged from $40,000 to $85,000, a price far beyond most modest new entrants. For this reason, FemtoSats arose with the WikiSat, [3] and PCBSat, [4], both small satellites with a mass below 100 grams. However, these new ideas presented have thermal insulation problems and are limited by current technologies. We have proposed similar platforms that would morph shape into spheres or attain wheels and body using inflatables and soft-body shells to overcome thermal concerns [8]. These first two types of FemtoSats propose a flat architecture that is very compact but unable to protect itself from extreme temperature conditions. In 2016, there was an initiative to create a standard for FemtoSats with the SunCube where the 1F has a mass of 35 grams and a 3F with a mass of 100 grams. This project adopts this cubic architecture from the SunCube standard. FemtoSats introduce whole new application thanks to their low-cost and disposability. This includes use of FemtoSat swarms for space monitoring, multipoint observation and tracking, in-addition to use of FemtoSats inside of science laboratories to demonstrate mobility on low-gravity environments [9, 10]. The small stowage size of FemtoSat can be a disadvantage for long-range communication which requires high gains. Inflatables and deployables can be used to deploy large antennas that address these needs [11].

In summary, the main challenge in space exploration is cost. Finding new and cost-effective ways to explore space, explore unknown environments such as the Permanently Shadowed Regions (PSR) on the Moon opens to the possibility to new mission concepts using low-cost disposable spacecraft with whole new modes of exploration and discovery. Small satellites have arisen thanks to miniaturization and COTS, however CubeSats costs are still too high for university teams, high school projects and individual initiatives. To overcome this accessibility problem ever smaller systems such as FemtoSats have appeared, however some designs such as a Chipsat or a Boardsat are more prone to thermal degradation and impose challenges in terms of pointing and 3-axis attitude control when needed. Having an enclosed volume as advocated by the SunCube standard better equips designer to deal with thermal considerations as here thermal design is a major challenge. For this mission, a 2F FemtoSat (30mm x 30 mm x 60 mm) is sought, with a mass below 100 grams and a manufacturing cost around $300.

## OBJECTIVE, MOTIVATION AND CHALLENGES

The aim of this paper is to develop a new FemtoSat conceptual design, low-cost and low mass for exploration inside a lunar permanently shadowed crater (Figure 1). In addition, we expect to show results on the parts that have been currently tested, to proof laser charging and radio localization. Finally, we aim that this project serves to get closer to PSR exploration.

This project is driven by the exploration of the unknown: study the Moon's PSRs and proof new concepts like laser charging and radio localization.

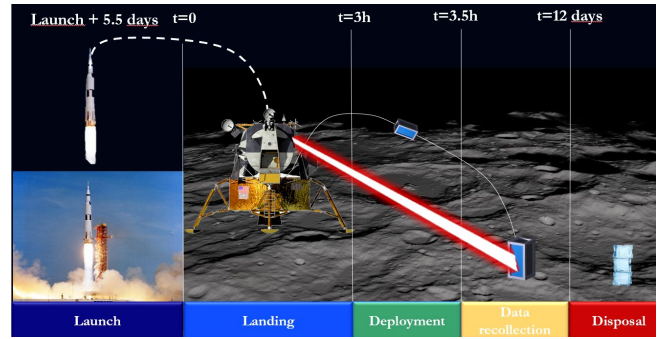

**Figure 1. Mission Concept of Operations using FemtoSats to explore lunar PSRs.**

For these goals to be achieved several problems must be overcome, such as, extreme temperatures and uncertainty. Temperatures range from -153ºC to 107 ºC and the biggest problem is that it is unknown what the conditions will be when the Lander gets to the Moon. Hence, the design must cover all possibilities. In addition, we yet do not know if the components are going to work in the given conditions. We ignore what the terrain is like, if it's sharp were the FemtoSats are going to land or if it is soft Lunar regolith.

## DESIGN

As mentioned before the intention is to design a 2F FemtoSat and its compounded of four major subsystems: structure, hardware, thermal, and power. For the structure, an aluminum box will contain the hardware and the insulation. In addition, a protector case, made of a high impact absorbent material, will contain the aluminum box and the solar panels. Finally, since the case is just a frame, gorilla glass will be added separating the solar panels from the outside.

The hardware will have a Tinyduino processor board or Tinyzero board, both from Tinycircuits, [5]. In addition, it will count with an IMU, a 433 MHz radio and a protoboard with pins. In order to protect the solar panels better it is desirable that they don't withstand any force in equilibrium. For this reason, the solar panels are made smaller than the aluminum box. This makes a frame for the solar panels of 21 mm x 51 mm in the long faces and 21mm x 21mm in the short faces. The aluminum box outer dimensions are 24.8 mm x 24.8 mm x 54.8 mm and have a thickness of 0.5 mm.

Regarding the thermal insulation, the intended MLI thickness is 0.5 mm and adding Aerogel is under consideration. The size of the hardware elements is 20 mm x 20 mm. This leaves almost 3 mm in the inside for wiring and other possible elements. On the power system side, we will use solar panels designed for space applications, and will account for the increase in current and voltage due to temperature.

This section is divided in three parts: the experiments or milestones, thermal analysis and communications. In addition, the experiments are three: height survival, power system and radio localization.

*Impact survival*

The first experiment, impact survival, is based on a conservation of energy model since no air resistance or friction occur between launching and landing in the FemtoSat's trajectory (Figure 2). FemtoSats are going to be thrown within a range of 2 to 20 meters from a height of 1.9 meters. They need to be designed so that they survive the impact with the ground, but in order to make the design, the energy that needs to be absorbed must be known.

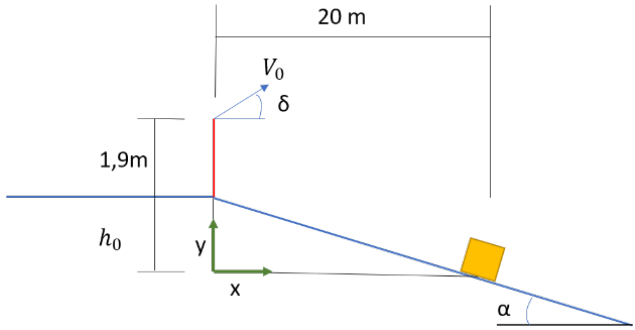

**Figure 2. Sketch for the energy model. The lander is represented in red and the FemtoSat is represented in yellow. α represents the mean slope between the lander and the FemtoSat and δ, the launching angle. $V_0$ represents the initial velocity of the FemtoSat and $h_0$ represents the vertical distance from the FemtoSat to the base of the lander.**

The system has one DOF and one parameter, $\delta$ and $\alpha$ respectively, hence an optimization can be done resulting in just one local minimum or maximum. In addition, since the number of equations and variables are the same, there is an analytical expression that calculates the launching angle, $\delta$ as a function of the parameter $\alpha$.

$$\sin 2\delta \,(20 \tan \alpha + 1.9 + 20 \tan \delta) = 20 \quad (1)$$

This expression provides, for this experiment, the launching angle that minimizes the impact energy, for this particular case. However, there is a general expression for similar types of experiments:

$$\sin 2\delta \,(x_o \tan \alpha + z_o + x_o \tan \delta) = x_o \quad (2)$$

Where $x_o$ and $z_o$ represent the desired range or the horizontal flying distance of the FemtoSat and the height measure from the base of the lander from where is launched (same as $h_0$), respectively.

*Solar array design*

The present part intends to provide an estimation on what $V_{MP}$ and $I_{MP}$ in standard conditions are needed for the solar panels DHV technologies provide. Some parameters have been estimated using typical orders of magnitude. The equations used for these calculations are:

$$I_{MP}^{STD} = \frac{\frac{\Phi}{\Phi^{STD}} I_{MP}}{n_1} - \frac{\delta J}{\delta T}(T - T_{STD}) A \quad (3)$$

$$V_{MP}^{STD} = \frac{V_{MP}}{n_2} - \frac{\delta V}{\delta T}(T - T_{STD}) \quad (4)$$

Where $n_1$ is the number of cells in parallel and $n_2$ is the number of cells in series. Results for just one cell are optimum since the solar cell could work in both hot and cold condition. The desired voltage, ranges between 4.64 and 4.99 V and the current would be 28 mA. However, there is not a cell with those capabilities in the market since voltage is too high. For that reason, there is a need of using at least two cells in series.

*Thermal design*

The thermal analysis provides an idea of the type of insulation needed. It shows if passive methods are enough or if the mission requires also the use of active methods to regulate the temperature. The advantages of the former are their simplicity and the lack of power consumption. On the other hand, the latter allows the designers to control the FemtoSat temperature precisely. For the thermal model the electric analogy is used. The system is composed of the hardware ($E_{bh}$), the MLI ($E_{bMLI}$), the aluminum box ($E_{bAl}$), the solar panels ($E_{bSP}$), the glass ($E_{bg}$), the Lunar surface ($E_{bsoil}$) and the outer space ($E_{b2}$). The thermal system is as follows

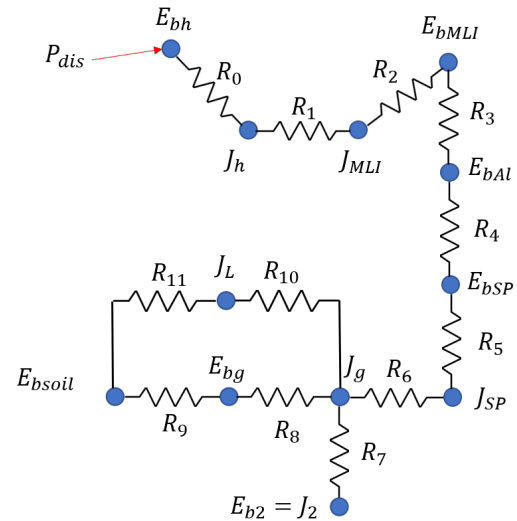

**Figure 3. Electrical analogy for the thermal analysis.**

$$R^i = \frac{1-\varepsilon_m}{A_m \varepsilon_m} \quad (5)$$

$$R^e = \frac{1}{A_h F_{p-q}} \quad (6)$$

$$R^c = \frac{t_m}{A_m k_m} \quad (7)$$

Where $R^i$ indicates the internal resistor of the element due to radiation, $\varepsilon_m$ indicates the emissivity of the material or element "m", $A_h$ represents the radiation emitting area and $F_{p-q}$ shows the sight facto between element "p" and element "q". $R^e$ indicates the radiation resistor between two elements connected, while $R^c$ represents the resistor due to conductivity. Finally, the thickness of element "m" is indicated by $t_m$ and the thermal conductivity coefficient by $k_m$.

**Table 1. Resistors values.**

| | |
|---|---|
| $R^i_0$ | 0 |
| $R^e_1$ | 376 |
| $R^i_2$ | 0 |
| $R^c_3$ | 0 |
| $R^c_4$ | $3.58 \cdot 10^{-4}$ |
| $R^i_5$ | 64.52 |
| $R^e_6$ | 193.57 |
| $R^e_7$ | 2436.2 |
| $R^i_8$ | 59.6 |
| $R^c_9$ | 0.47 |
| $R^e_{10}$ | 1102.3 |
| $R^i_{11}$ | 0 |

The equation for the resistors calculations can be found in [Appendices]. Properties of the materials used can also be found on the [Appendices]

Finally, to calculate $P_{dis}$, which stands for dissipated power, we need to analyze what is the outpower of the battery and what is the power consumption of the elements.

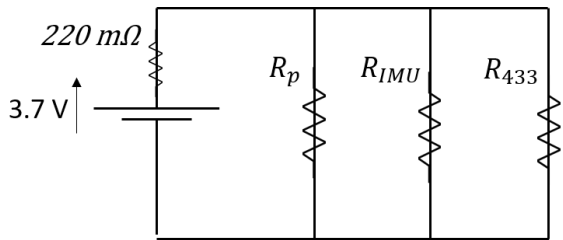

**Figure 4. Circuit analysis**

From Figure 4, the voltage and current that goes through each element enables us to calculate the dissipated power.

$$P_{dis} = P^{bat}_{dis} + P^{proc}_{dis} + P^{IMU}_{dis} + P^{433}_{dis} \quad (8)$$

$$P_{dis} = 10.99 \ mW$$

Now that all the numeric values are obtained, it is just a matter of solving the circuit in Figure 3. There are two temperatures that are important to obtain: the hardware temperature and the solar panels temperature.

$$\frac{E_{bMLI} - E_{bh}}{\Sigma^2_{i=0} R_i} + P_{dis} = 0 \quad (9)$$

$$P_{dis} = \frac{E_{bMLI} - E_{bSP}}{R_3 + R_4} = \frac{T_{MLI} - T_{SP}}{\frac{t_{MLI}}{A_{MLI} k_{MLI}} + \frac{t_{Al}}{A_{Al} k_{Al}}} \quad (9)$$

$$E_{bSP} \left( \frac{1}{R^*} + \frac{1}{R_7} \right) = \left( \frac{E_{bsoil}}{R^*} + \frac{E_{b2}}{R_7} \right) + P_{dis} \left[ 1 + (R_5 + R_6) \left( \frac{1}{R^*} + \frac{1}{R_7} \right) \right] \quad (10)$$

$$E_{bi} = \sigma T_i^4 \text{ (for radiation)} \quad (11)$$

Where $\sigma$ is Boltzmann's constant. The system although it seems coupled can be solved easily in the order (10), (9), (8). In addition, the outer space temperature can be considered to be 0 K, thus $E_{b2}$ is zero.

Since we are designing for cold conditions the temperature of the soil is -153 ºC (122.15K). Note that for this section temperatures area treated in Kelvin.

*Radio frequency localization*

The great difficulty of this section is that it is not possible to use GPS tracking as it is done on Earth. If there was a satellite constellation providing GPS localization it would be much easier. However, this problem may be possible to overcome by using the Decawave DWM 1000 [Decawave] (Figure 5). This device allows 10 cm localization precision. Power consumption is in accordance with the rest of the hardware, it consumes between 2.8 to 3.6 V and 31 mA when receiving and 61 mA when transmitting. Dimensions of this device also fits in the 2F FemtoSat (23 mm x 13 mm x 2.9 mm). Finally, the temperature range is also in accordance with the rest of devices (-30ºC to 85ºC).

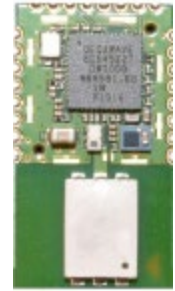

**Figure 5. DWM 1000 device from Decawave. Source [6].**

*Circuitry*

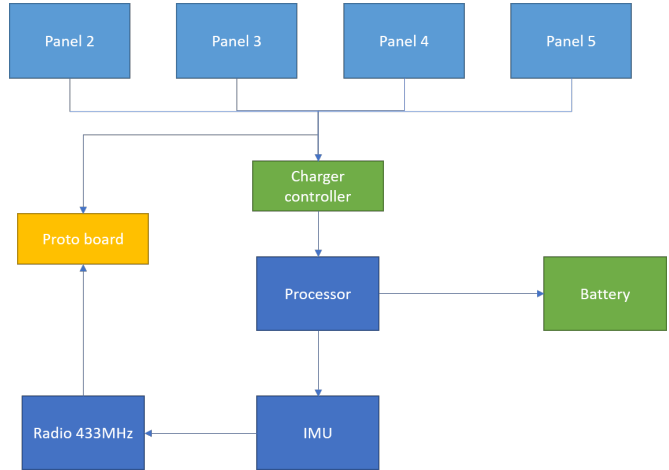

**Figure 6. System Architecture design.**

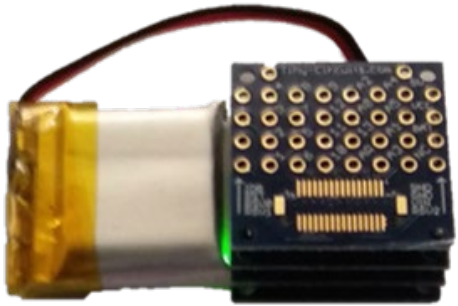

**Figure 7: Assembled system boards.**

The overall architecture and packed FemtoSat shown in Figure 6 and 7. These results were validated with two numerical analysis. The former implemented an optimization and the second one just calculated the energy for every combination of $\delta$ and $\alpha$. With these results now it is easy to compute the energy per mass unit and the equivalent height on Earth for testing,

$$H_E = \frac{E_m}{mg_E} \cdot SM = 2.18\ m$$

Where $H_E$ is the height on Earth, $E_m/m$ is the mechanical energy per mass unit, $g_E$ is Earth's gravity and SM is the safety margin and it has been considered 10%. Figure 8 represents the variation of the mechanical energy with the mean slope and confirms that most critic case is the one with bigger slope.

## RESULTS

This section will gather the results obtained first for the design part and then what measurements were obtained from the experiments.

*Impact survival*

**Table 2. Impact results for a 12º mean slope.**

| | |
|---|---|
| $v_0\ (m/s)$ | 4.97 |
| $t\ (s)$ | 5.01 |
| $\delta\ (º)$ | 36.45 |
| $h_0\ (m)$ | 4.25 |

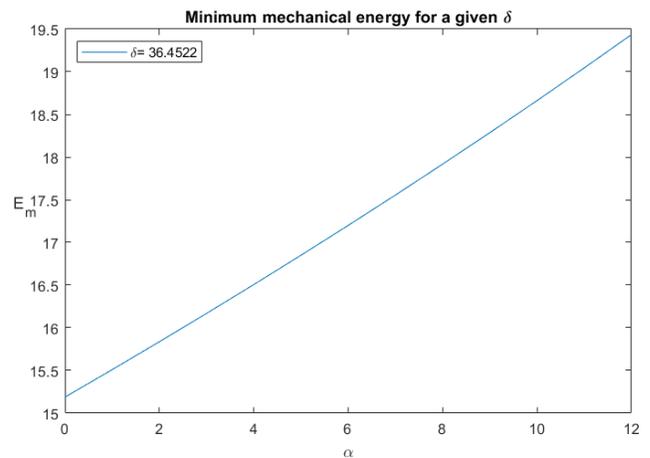

**Figure 8. Representation of the minimum mechanical energy for each α at δ= 36.54.**

*Solar array*

Results with four cells, two in parallel and two in series are good, but still present some problems. It was discovered that the greater the number of cells the more distortion between hot and cold conditions exist, therefore it is interesting to minimize the number of cells in series.

| Φ | T | Imp | Vmp |
|---|---|---|---|
| 1322 | 107 | 72.75 | 3.1 |
| 4444 | 107 | 244.59 | 3.1 |
| 4444 | -153 | 81.63 | 6.01 |
| 1414 | 107 | 77.82 | 3.1 |

**Table 3. Maximum power current and voltage for different radiation and temperature.**

From Table 3, it can derived that if we design for warm conditions and we land in cold conditions the charger controller and therefore the battery will burn. On the other hand, if we design for cold conditions and land in warm conditions the batteries will charge up to a certain percentage, small. So far let's just focus on solving the voltage problem. It would be a great advance if we could have a $V_{MP}$ of almost 6V in cold conditions and at least obtain 3.7V in warm conditions, which is the nominal value of the battery and would allow the battery to charge up to a decent percentage. If we get to raise the temperature of the panels to -98ºC in cold conditions, we get the desired condition.

*Thermal analysis*

Using a standard MLI and using all the thickness available, 2.3 mm, with a thermal conductivity coefficient of 0.01 the dissipated power to achieve the temperature requirements is 290 mW, achieving a temperature on the equipment and the panels respectively of,

$$T_{SP} = -67 \text{ ºC}$$

$$T_h = -19\text{ºC}$$

These results show no efficiency at all, since we would like to have the solar panels around -95ºC. In addition, the battery lasts in these conditions 1.5h. Solutions to improve this are decreasing the thermal conductivity of the MLI and probably combining it with aerogel. The optimum result could be achieved by having a thermal conductivity coefficient of 0.001 which is not that far from current results and may be doable by combining aerogel and MLI.

$$T_h = -18 \text{ ºC}$$

$$T_{SP} = -93\text{ºC}$$

In this optimum result the dissipated power required is 155 mW and if using a 150 mAh the working time is 3h. If changing the battery to one with 500 mAh the science life would be 10h.

*Mass budget*

**Table 4. Mass and volume of the different elements in the FemtoSat.**

| Group | Element | Mass (g) | Volume (mm³) |
|---|---|---|---|
| Hardware | Processor board (TinyZero) | 1.4 | 1160 |
| | IMU | 1 | 2044 |
| | Radio + Antenna | 1.41 | 2044 + (177 mm) |
| | ProtoBoard | 0.85 | 2044 |
| Power system | Battery (500 mA, 3.7V) | 9.3 | 5890 |
| | Charger Controller | 0.54 | 466.2 |
| | Solar Panels 4x | ~27 | 6854 |
| Structure | TPU case | 9 | 54000 |
| | Box (aluminum) | 9 | 33704 (t=0.5 mm) |
| Insulation | MLI | TBD | 12509 |
| | Aerogel | ~0 | TBD |
| *Total* | | 54 | 54000 |

*Power budget*

**Table 5. Power budget of the different elements and Temperature range.**

| Group | Element | Tempeature Range (ºC) | Current (mA) | Voltage (V) | Power (mW) |
|---|---|---|---|---|---|
| Hardware | Processor board (TinyZero) | [-40, 85] | 4 | 2.7 – 5.5 | 10.8 – 22 |
| | IMU | [-40, 85] | 4.6 | 3.5 – 5 | 16.1 – 23 |
| | Radio + Antenna | [-40, 85] | 18.5 (RX)<br>30 (TX 13 dBm)<br>85 (TX 20 dBm) | 3.5 – 5 | 65 – 93<br>105 – 150<br>298 - 425 |
| | ProtoBoard | Unspecified | | 3.7 | TBD |
| Power system | Battery (500 mA, 3.7V) | [-20, 130] | 30-500 | | 1.85 (Wh) |
| | Charger controller | [-40, 85] | 15-500 | 4.2 – 6 | |
| | Solar Panels | Unspecified | 65 | 4.66 | 303 |

## DISCUSSION

Outer space missions are complex and therefore a lot remains to study about FemtoSats but some relations can be drawn. If FemtoSats are to be ejected from a lander it is imperative that we design a trajectory that minimizes the impact energy since equipment is sensitive. In the same line they need thermal protection and thus studying and selecting the proper materials is important. For this reason, maybe it is a better option to substitute the aluminum box for a different material with lower thermal conductivity coefficient.

In this paper it has been shown that the power system design is coupled with the thermal design. For that reason, it is recommendable to make a second order analysis or use finite element methods to have more accurate results. Moreover, this gains importance since every minor change could make the difference between having available volume or not.

Another important aspect to notice is that even though here on Earth it is possible to perform testing with commercial solar panels in outer space those components wouldn't work since they normally have multiple cells in series and therefore increasing the sensitivity of the voltage due to temperature until a point where the charger controller cannot handle.

Transmitting and receiving information consumes a great amount of power, hence their duty cycle is going to be minimized when experiments are carried out.

First of all, for a given maximum mean slope of 12º provided by Colorado School of Mines from his project LASER, [7], the ejecting angle that minimizes the mechanical energy is 36.45º in order to reach 20 meters horizontally. This means that the initial speed at which the FemtoSat is launched is almost 5 m/s and therefore the system that launches the FemtoSat will have to be designed in accordance with this value. In order to test if the solar cells and the hardware inside would survive, the FemtoSat needs to be dropped from a 2 m height and accounting for 10% safety margin it is recommended to drop it from 2.2 m.

In order to be able to charge the solar panels in both hot and cold conditions we need two cells in series per solar panel so that the final output voltage at maximum power in standard conditions is 3.98 V. At the same time an insulation system that can keep the solar panels at a temperature of 95 ºC or less is required. This also allows to dissipate less power to warm hardware and battery. Finally, radio power consumption is high therefore duty cycle should be minimum.

## CONCLUSIONS

Overall, we show it is feasible for us to deploy FemtoSats from the lander into the PSR to provide exploration data. For the FemtoSat deployment problem, the ejecting energy should be calculated for the maximum mean slope expected. In addition, the problem just presents one degree of freedom therefore an analytic solution can be found. Current commercial solar panels are too sensitive to temperature for that reason it is not recommended to use them in space if our battery charging voltage is limited. To avoid this problem the main solution is to minimize the number of cells in series. Power analysis and thermal analysis are highly coupled, therefore higher order analysis or finite element methods should be used. However, the linear method used here provides a quick insight of the requirements. In order to maximize the time the mission lasts the radio duty cycle should be minimized since its power consumption is very high.

# APPENDICES

## A. Thermal properties

**Table 6. Different materials emissivities.**

| Material | Emisivity |
|---|---|
| Aluminum | 0.04 |
| Glass | 0.94 |
| Coper | |
| White paint | 0.95 |
| Black paint | 0.97 |
| Solar panel | 0.75 |

**Table 7. Different elements thermal conductivity coefficient, k.**

| Material | Conductivity |
|---|---|
| Aluminum | 209.3 |
| Glass | 0.8 |
| MLI | 0.01 |

**Table 8. Sight factor between components.**

| Components | Sight Factor |
|---|---|
| $F_{h-MLI}$ | 1 |
| $F_{SP-g}$ | 1 |
| $F_{g-soil}$ | 0.3 |
| $F_{g-2}$ | $1 - F_{g-soil}$ |

**Table 9. Components area (mm²).**

| Component | Area (mm²) |
|---|---|
| $A_h$ | 2658.4 |
| $A_{MLI}$ | 6254.64 |
| $A_{Al}$ | 6666.24 |
| $A_{SP}$ | 5166 |
| $A_{g1}$ | 1071 |
| $A_{g2}$ | 3025 |
| $A_{soil}$ | ∞ |

## B. Resistors detailed calculation

**Table 10. Internal and external radiation resistor and thermal conductivity resistor.**

| | |
|---|---|
| $R_0^i$ | $\dfrac{1-\varepsilon_h}{A_h \varepsilon_h} = 0$ |
| $R_1^e$ | $\dfrac{1}{A_h F_{h-MLI}} = \dfrac{1}{A_h} = 376$ |
| $R_2^i$ | $\dfrac{1-\varepsilon_{MLI}}{A_{MLI}\varepsilon_{MLI}} = 0$ |
| $R_3^c$ | $\dfrac{t_{MLI}}{A_{MLI} k_{MLI}} = 0$ |
| $R_4^c$ | $\dfrac{t_{Al}}{A_{Al} k_{Al}} = 3.58 \cdot 10^{-4}$ |
| $R_5^i$ | $\dfrac{1-\varepsilon_{SP}}{A_{SP}\varepsilon_{SP}} = 64.52$ |
| $R_6^e$ | $\dfrac{1}{A_{SP} F_{SP-g}} = 193.57$ |
| $R_7^e$ | $\dfrac{1}{A_{g1} F_{g-2}} + \dfrac{1}{A_{g2} F_{g-soil}} = 2436.2$ |
| $R_8^i$ | $\dfrac{1-\varepsilon_g}{A_g \varepsilon_g} = 59.6$ |
| $R_9^c$ | $\dfrac{t_g}{A_g k_g} = 0.47$ |
| $R_{10}^e$ | $\dfrac{1}{A_{g2} F_{g-soil}} = 1102.3$ |
| $R_{11}^i$ | $\dfrac{1-\varepsilon_{soil}}{A_{soil}\varepsilon_{soil}} = 0$ |

## C. Hardware dissipated power calculation

$$P_{dis} = P_{dis}^{bat} + P_{dis}^{proc} + P_{dis}^{IMU} + P_{dis}^{433}$$

$$P_{dis} = I^2 R_{bat} + V_{proc} I_{proc} + V_{IMU} I_{IMU} + V_{433} I_{433}$$

$$P_{dis} = 27.1 * 0.027 * 0.2 + 0.4 * 4 + 0.4 * 4.6 + 0.4 * 18.5 = 10.99 \; mW$$

$$\frac{1}{R^*} = \frac{1}{R_8 + R_9} + \frac{1}{R_{10}}$$

## ACKNOWLEDGEMENT

Thank you to Team LASER from Colorado School of Mines to offer us the opportunity to work on this project and participate with them in NASA Big Idea Challenge 202

## BIOGRAPHY

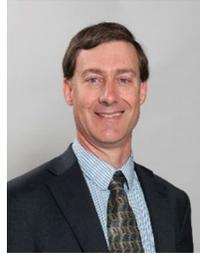

***Leonard Dean Vance*** *is a PhD candidate in Aerospace and Mechanical Engineering at the University of Arizona, and a retired Senior Fellow from Raytheon Missile Systems. His 33 year career at Hughes Aircraft and Raytheon Missile Systems includes Systems Engineering lead, Chief Engineer and Program manager for a variety of forward leaning projects, including the LEAP kinetic kill vehicle, the AIM-9X sidewinder missile, Kinetic Energy Interceptor and program manager for the SeeMe imaging microsatellite. He holds a masters of Engineering from Harvey Mudd College, six patents and is currently in his 3rd year of the PhD. Program at the University of Arizona.*

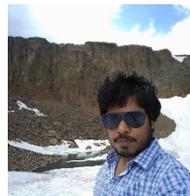

***Himangshu Kalita*** *received a B.Tech. in Mechanical Engineering from National Institute of Technology, Silchar, India in 2012. He is presently pursuing his Ph.D. in Mechanical Engineering from the University of Arizona in the Space and Terrestrial Robotic Exploration (SpaceTREx) Laboratory. His research interests include dynamics and control, space robotics, machine learning and automated design.*

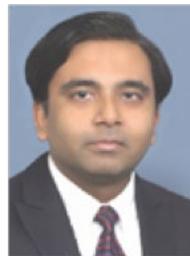

***Jekanthan Thangavelautham*** *has a background in aerospace engineering from the University of Toronto. He worked on Canadarm, Canadarm 2 and the DARPA Orbital Express missions at MDA Space Missions. Jekan obtained his Ph.D. in space robotics at the University of Toronto Institute for Aerospace Studies (UTIAS) and dis his postdoctoral training at MIT's Field and Space Robotics Laboratory (FSRL). Jekan Thanga is an assistant professor and heads the Space and Terrestrial Robotic Exploration (SpaceTREx) Laboratory at the University of Arizona. He is the Engineering Principal Investigator on the AOSAT I CubeSat Centrifuge mission and is a Co-Investigator on SWIMSat, an Airforce CubeSat mission concept to monitor space threats.*